\documentclass[a4paper,prl,twocolumn,showpacs]{revtex4}

\usepackage{amssymb}
\usepackage{amsfonts}
\usepackage{amsmath}
\usepackage{euler}
\usepackage{color}
\usepackage{graphics, graphicx}

\newcommand{\ff}[1]{{\boldsymbol #1}}

\begin{document}

\author{Andrej Schwabe, Irakli Titvinidze and Michael Potthoff}

\affiliation{I. Institut f\"ur Theoretische Physik, Universit\"at Hamburg, Jungiusstra\ss{}e 9, 20355 Hamburg, Germany }


%71.70.Gm	Exchange interactions
%75.10.Lp	Band and itinerant models
%75.75.-c 	Magnetic properties of nanostructures
%85.75.-d 	Magnetoelectronics; spintronics: devices exploiting spin polarized transport or integrated magnetic fields

\title{Inverse indirect magnetic exchange}

\begin{abstract}
Magnetic moments strongly coupled to the spins of conduction electrons in a nanostructure can confine the conduction-electron motion due to scattering at almost localized Kondo singlets. 
We study the resulting local-moment formation in the conduction-electron system and the magnetic exchange coupling mediated by the Kondo singlets. 
Its distance dependence is oscillatory and induces robust ferro- or antiferromagnetic order in multi-impurity systems.
\end{abstract}

\maketitle

%--------------------------------------------------------------------------------------------------------------------------------------
\paragraph{Introduction.}

The appearance of magnetic order in condensed-matter systems 
\cite{Yos96,NR09}
requires 
(i) the existence or the formation of local magnetic moments,
(ii) a coupling mechanism favoring a certain alignment of the moments, e.g.\ ferro- or antiferromagnetically,
and 
(iii) the stability of long-range magnetic order against different types of thermal or quantum fluctuations and against competing ordering phenomena.

Local-moment formation typically results from incompletely filled localized orbitals or from strong local correlations and can be described by Hubbard-, Anderson- or Kondo-type models \cite{And61,Hub63,Kon64}.
Among the different known coupling mechanisms, such as the direct Heisenberg exchange \cite{hei28} or the indirect Anderson super exchange \cite{Kra34,And50}, the Ruderman-Kittel-Kasuya-Yoshida (RKKY) interaction \cite{RK54,Kas56,Yos57} provides a mechanism for longer-ranged coupling $J_{\rm RKKY} \sim \pm 1/r^{D}$ between magnetic impurities in a $D$-dimensional metallic system which can be either ferro- or antiferromagnetic, depending on the distance.
It originates from a local exchange coupling $J$ which, for weak $J$, mediates an indirect interaction $J_{\rm RKKY} \propto J^{2}$.

RKKY exchange has gained much interest recently in the context of nanostructures, e.g.\ in double-dot semiconductor quantum devices \cite{CTL+04} with tunable RKKY-mediated control on spin degrees of freedom. 
Nanostructures with tailored magnetic properties can be engineered using scanning-tunnelling techniques by positioning magnetic atoms on non-magnetic metallic surfaces at certain distances where $J_{\rm RKKY}$ is ferro- or antiferromagnetic \cite{ZWL+10,KWCW11,KWC+12}.
Furthermore, RKKY-mediated magnetism competes \cite{Don77} with the Kondo effect \cite{Kon64,hew93}, i.e., the screening of the local magnetic moments due to non-local antiferromagnetic correlations induced by $J$. 
A subtle Kondo-vs.-RKKY competition takes place at weak $J$ in nano-systems with strong electron-confinement effects  \cite{TKvD99,SGP12}.

Here, we study an exchange mechanism where the roles of conduction electrons and impurities are ``inverted''.
We show that the Kondo effect {\em helps} (i) to form local moments, (ii) to couple the moments and (iii) leads to magnetic order in certain nano-structured geometries:
For strong $J$, almost local Kondo singlets are formed which act as hard scattering centers for the itinerant conduction electrons and may confine their motion, depending on the impurity positions.
In certain geometries, this tends to localize the conduction electrons and leads to the formation of local magnetic moments in the {\em a priori} uncorrelated conduction-electron system.
These moments are found to couple magnetically via virtual excitations of the Kondo singlets.

We study the resulting ``inverse indirect magnetic exchange'' (IIME) by means of strong-coupling perturbation theory and different numerical techniques. 
The IIME shows an oscillatory distance dependence.
For extended systems, it triggers long-range magnetic order which is robust against charge fluctuations on the impurities but sensitively depends on the quantum confinement of the conduction electrons, e.g., on the geometry of magnetic adatoms in an experimental setup using scanning-tunnelling techniques.

Furthermore, for certain geometries, the IIME at strong $J$ can be understood as evolving by adiabatic connection from the standard RKKY coupling at weak $J$. 
This is ensured by quantum confinement and by exact results \cite{She96,Tsu97a} based on Lieb's concept of reflection positivity in spin space \cite{Lie89} available for Kondo systems on bipartite lattices at half-filling.

%--------------------------------------------------------------------------------------------------------------------------------------
\paragraph{From RKKY to inverse exchange.}

We consider a system with $R$ spins $\ff S_{r}$, with spin-quantum numbers $1/2$, which are coupled locally via an antiferromagnetic exchange $J>0$ to the local spins $\ff s_{i}$ of a system of $N$ itinerant and non-interacting conduction electrons. 
The conduction electrons hop with amplitude $t\equiv 1$ between non-degenerate orbitals on neighboring sites of a $D$-dimensional lattice of $L$ sites:
\begin{equation}
{\cal H} = - t \sum_{\langle i,j \rangle, \sigma} c^{\dagger}_{i\sigma} c_{j\sigma} + J \sum_{r=1}^{R} \ff s_{i_{r}} \ff S_{r} \: .
\label{eq:ham}
\end{equation}
Here, $c_{i\sigma}$ annihilates an electron at site $i=1,...,L$ with spin projection $\sigma=\uparrow, \downarrow$, and 
$\ff s_{i} = \frac{1}{2} \sum_{\sigma \sigma'} c^{\dagger}_{i\sigma} \ff \sigma_{\sigma\sigma'} c_{i\sigma'}$ is the local conduction-electron spin at $i$, where $\ff \sigma$ is the vector of Pauli matrices.
Impurity spins couple to the local conduction-electron spins at the sites $i_{r}$.
We investigate the half-filled system with $N=L$ electrons.

\begin{figure}[b]
\centerline{\includegraphics[width=0.4\textwidth]{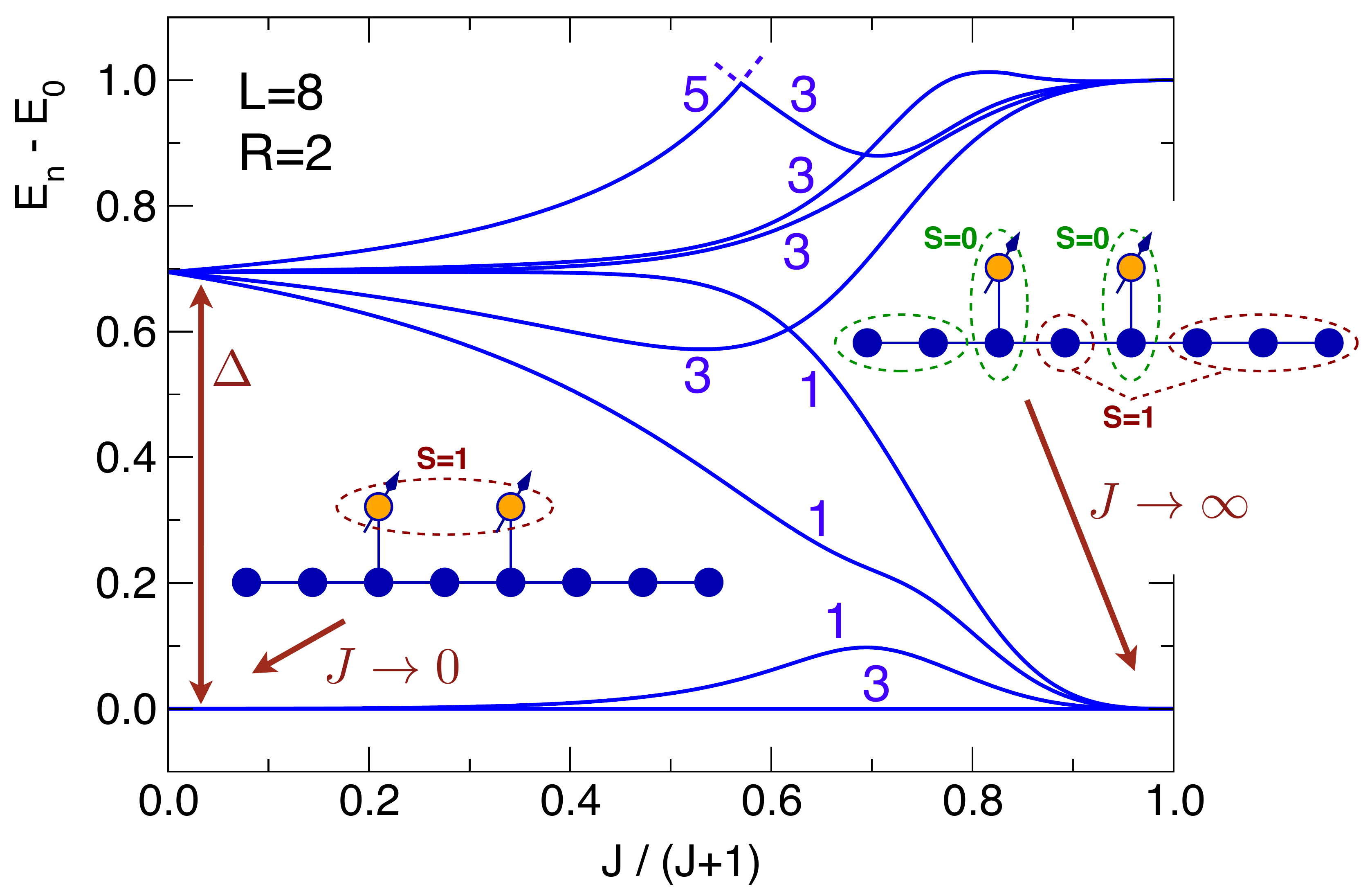}}
\caption{(Color online) 
Inverse indirect magnetic exchange (IIME) mechanism.
Lowest many-body eigenenergies $E_{n}-E_{0}$ for $n=0,...,7$ in the entire range $0 < J < \infty$ (note the nonlinear scale) for the model (\ref{eq:ham}) with $L=8$ host sites and $R=2$ impurity spins 1/2 at the sites $i_{1}=3$ and $i_{2}=5$. 
Multiplicities are indicated (see numbers).
The system smoothly crosses over from conventional indirect RKKY exchange ($J\to 0$) to a state for $J\to \infty$ where two local Kondo singlets lead to the formation of two spins $1/2$ in the host which are coupled to a triplet {\em via} a magnetically inert Kondo singlet.
Pictograms: dominant spin correlations obtained numerically.
Energy scale set by the n.n.\ hopping $t=1$.
}
\label{fig:gaps}
\end{figure}

To illustrate the crossover from conventional RKKY indirect magnetic exchange at weak $J$ to the inverse indirect exchange at strong $J$, we first analyze a simple model with a small number of $L=8$ sites and $R=2$ spins at $i_{1}=3$ and $i_{2}=5$ using exact diagonalization (see Fig.\ \ref{fig:gaps}).
In the RKKY regime for $J\to 0$, the low-energy sector of ${\cal H}$ is exactly described by an effective RKKY two-spin model $H_{\rm RKKY} = - J_{12} \ff S_1 \ff S_2$ with $J_{\rm 12} \propto (-1)^{|i_{1}-i_{2}|} J^2 / |i_{1}-i_{2}|$. 
For a ``ferromagnetic distance'' $i_{1}-i_{2}=2$ the two impurity spins form a non-local triplet in the ground state.

%----------------------------------------------------------------------------------------------------
%----------------------------------------------------------------------------------------------------

As is seen in Fig.\ \ref{fig:gaps}, the ground state is unique (apart from the spin degeneracy) for any $J\ne 0$, $J\ne \infty$.
The absence of a ground-state level crossing at half-filling and for a bipartite lattice is in fact enforced by analytical results \cite{She96,Tsu97a}.
Consequently, the ground-state symmetry is preserved, and the triplet stays intact in the entire $J$ range. 
However, its character must change.
With increasing $J$, the Kondo effect, which for $J\to 0$ is cut by the finite-size gap $\Delta$, sets in 
and dominates for $J\to \infty$ with a shrinking and eventually local screening cloud.
This results in two completely local and magnetically inert ``Kondo'' singlets while the ground state must be a triplet. 
The numerical analysis of spin-correlation functions shows that this triplet is formed by two spins $1/2$ formed in different parts of the {\em conduction-electron} system which couple ferromagnetically. 
Opposed to a coupling of impurity spins mediated by the metal host for weak $J$, i.e.\ standard indirect RKKY exchange, this type of interaction represents an ``inverse indirect magnetic exchange'' (IIME) mediated by local Kondo singlets (see insets in Fig.\ \ref{fig:gaps}).

%--------------------------------------------------------------------------------------------------------------------------------------
\paragraph{Magnetic order.}

Before we analyze this exchange mechanism in detail, we demonstrate its usefulness to understand ferromagnetic order for large systems $L, R \to \infty$.
To this end we applied the density-matrix renormalization group (DMRG) \cite{Whi92,Sch11} to study one-dimensional models for different $L$ and for $R=(L+1)/2$ spins ($R$ odd) coupled to the sites $i_{r}=1,3,...,L-2,L$ (B sites, see Fig.2, inset), i.e.\ we study a ``diluted'' Kondo lattice with impurity spins at ``ferromagnetic'' distances.
At half-filling this model is known \cite{She96} to exhibit a ferromagnetic ground state.

Our implementation (see Ref.\ \cite{TSRP12} for details) makes use of conservation of the $z$-component of the total spin $\ff S_{\rm tot}=\sum_{r=1}^{R} \ff S_{r} + \sum_{i=1}^{L} \ff s_{i}$. 
The total spin $S_{\rm tot,0}$ is obtained by computing the ground-state expectation value 
$\langle {{\ff S}^{2}_{\rm tot}} \rangle = S_{\rm tot,0}(S_{\rm tot,0}+1)$.
For the system shown in Fig.\ 2 ($L=49$, $R=25$), we in fact find a large $S_{\rm tot,0} = (R-1)/2$ in the entire $J$ range. 
This is consistent with the prediction by Shen \cite{She96}.

The adiabatic connection between RKKY and IIME is more subtle in this case:
For $J\to 0$, standard RKKY theory would predict $S_{\rm tot,0}=S_{\rm max}=R/2$.
In the present case, exactly one of the impurity spins, however, is Kondo screened by the single electron occupying the spin-degenerate one-particle energy level at the Fermi wave vector of the non-interacting conduction-band system (see Ref.\ \cite{SGP12}). 
This results in $S_{\rm tot,0}=(R-1)/2$.
For all (finite but large) systems studied here, the ground state turns out to be a smooth function of $J$.
Therefore, $S_{\rm tot,0}=(R-1)/2$ must be the same in both limits. 
For $J\to \infty$, this large spin must then result from a ferromagnetic coupling of local magnetic moments at the sites $i=2,4, ..., L-1$ (A-sites, see Fig.2, inset) which are formed as a result of the increasing confinement of electrons due to the formation of local Kondo singlets at the B sites.
The DMRG calculations indeed yield strong antiferromagnetic local spin correlations $\langle \ff s_{i_{r}} \ff S_{r} \rangle \to -3/4$, vanishing RKKY correlations $\langle \ff S_{r} \ff S_{r'} \rangle \to0$, and local-moment formation $\langle \ff s_{i}^{2} \rangle \to 3/4$ at A-sites for $J\to \infty$.

\begin{figure}[t]
\centerline{\includegraphics[width=0.45\textwidth]{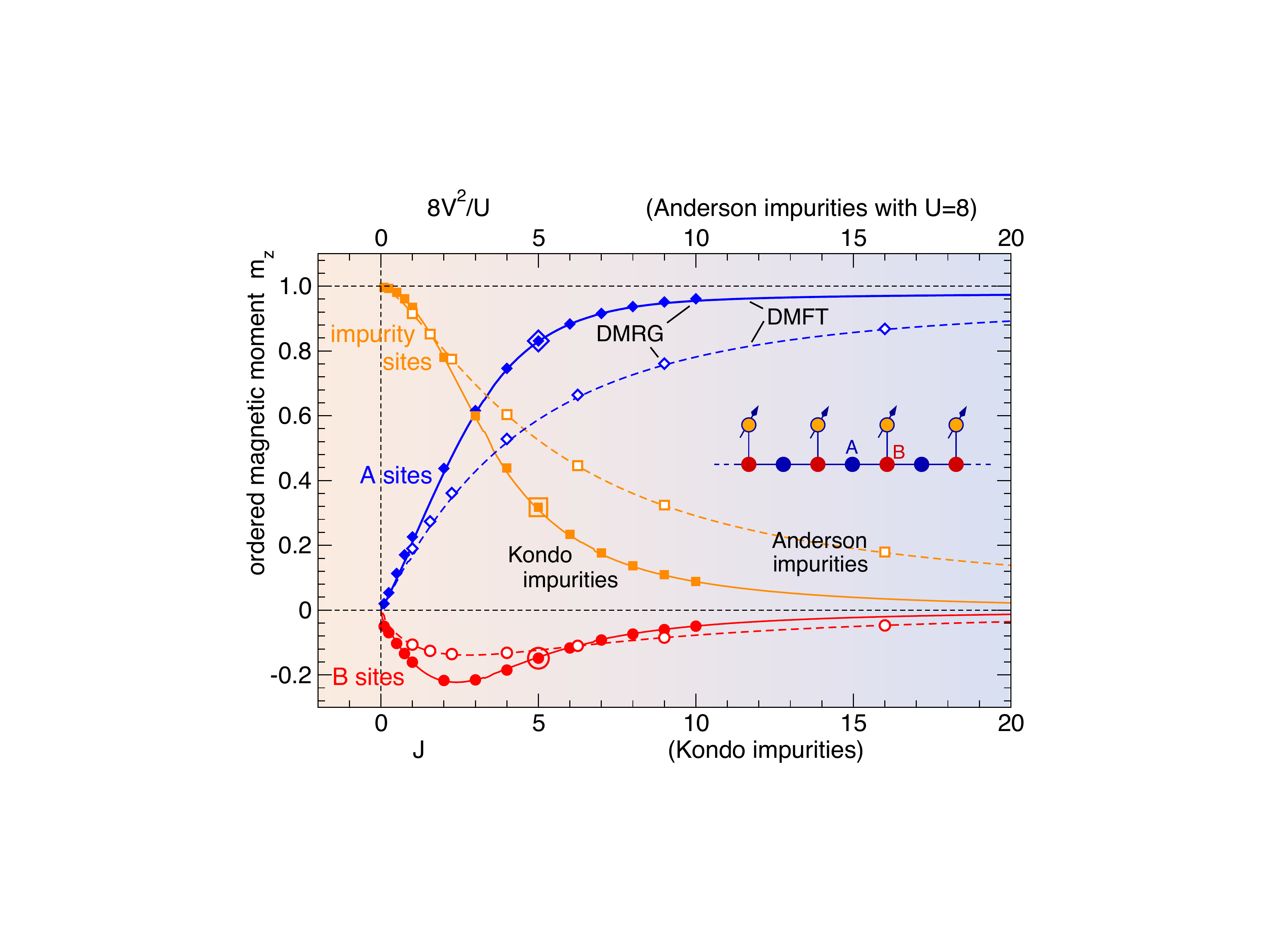}}
\caption{(Color online)
Crossover from the RKKY regime at weak coupling to the IIME regime at strong coupling. 
Calculated ordered magnetic moments on different sites of a tight-binding chain with spin-1/2 Kondo impurities as functions of J (solid lines, filled symbols) and with Anderson impurities as functions of $8V^{2}/U$ at Hubbard $U=8$ (dashed, open) -- see pictogram for system geometry.
Symbols: Density-matrix renormalization group (DMRG) for a system with $L=49$ uncorrelated sites (A and B), $R=25$ impurities (large symbols at $J=5$: $L=89$, $R=45$), open boundary conditions.
Lines: Dynamical mean-field theory (DMFT) for $L=100$, $R=50$, periodic boundary conditions.
Energy scale: $t=1$.
}
\label{fig:mag}
\end{figure}

Fig.2 shows the ordered magnetic moments at the central impurity $m_{\rm imp} \equiv 2 \langle S_{r,z} \rangle$, at the ``sub-impurity'' B site $m_{\rm B} \equiv \langle n_{i_{r}\uparrow} - n_{i_{r}\downarrow} \rangle$ and a neighboring A-site $m_{\rm A} \equiv \langle n_{i\uparrow} - n_{i\downarrow} \rangle$, as obtained from the ground state with maximum $M_{\rm tot} = S_{\rm tot,0}$.
With increasing $J$, there is a clear crossover from the RKKY regime, with $m_{\rm imp} \to 1$, $m_{\rm A},m_{\rm B} \to 0$, to the IIME regime for $J\to \infty$, where the magnetization of the system results from ordered moments at A-sites.
The results are characteristic for the infinite system as is obvious by comparing results for $L=49$ and $L=89$ (see $J=5$ in Fig.\ \ref{fig:mag}).

%--------------------------------------------------------------------------------------------------------------------------------------
%--------------------------------------------------------------------------------------------------------------------------------------

%--------------------------------------------------------------------------------------------------------------------------------------
\paragraph{Low-energy model.}

To analyze the mechanism generating a ferromagnetic coupling between magnetic moments at next-nearest neighboring A-sites $i$ and $j$, we treat the hopping term $\propto t$ in Eq.\ (\ref{eq:ham}) perturbatively. 
The starting point is the highly degenerate ground state of the $t=0$ model consisting of local Kondo singlets and an arbitrary electron configuration.
A non-trivial effective model capturing the low-energy sector of $H$ in the limit $0 < t \ll J$ is obtained at fourth order in $t$ through processes where e.g.\ an electron hops from $i\in A$ via the neighboring B site to $j\in A$ and, again via B, back to $i$. 
Here, the local Kondo singlet at B must be excited at an energy cost $\propto J$ first and restored again on the way back.
Calculations are lengthy but straightforward and will be published elsewhere \cite{TSP13}.
For $J>0$ and keeping terms up to ${\cal O}(t^{4}/J^{3})$ we find:
\begin{eqnarray}
  H_{\rm eff} / \alpha
  =
  - \sum_{i<j\in A} (\ff s_{i} \ff s_{j} - \ff t_{i} \ff t_{j} )
  + \sum_{i\in A} (n_{i\uparrow}-\frac{1}{2}) (n_{i\downarrow}-\frac{1}{2})
  \nonumber \\ 
  - \frac{1}{2}
  \sum_{i<j\in A} \sum_{\sigma} (c_{i\sigma}^{\dagger} c_{j\sigma}+\mbox{H.c.}) (1 - n_{i-\sigma} - n_{j-\sigma})
  \: .
\label{eq:eff}  
\end{eqnarray}
The effective model is governed by a single energy scale $\alpha \equiv 64 t^{4} / 3 J^{3}$ and describes spin and charge degrees of freedom on the A sites only. 
This is opposed to a strong-coupling variant of the RKKY theory \cite{TDH+13} where the focus is on the effective coupling between impurity spins. 

The first term in Eq.\ (\ref{eq:eff}) represents a Heisenberg-type ferromagnetic spin interaction and indeed explains the ferromagnetic IIME through a local Kondo singlet.
Ferromagnetism due to the IIME competes with formation of a charge-density wave or $\eta$ pairing \cite{TSU97b} as favored by the second term.
This includes the local isospin
$\ff t_{i} = \frac{1}{2} (c^{\dagger}_{i\uparrow}, (-1)^{i} c_{i\downarrow}) 
\cdot \ff \sigma  \cdot 
(c_{i\uparrow} , (-1)^{i} c^{\dagger}_{i\downarrow})^{T}$. 
Note that the total isospin $\ff T_{\rm tot}=\sum_{i} \ff t_{i}$ and the total spin $\ff S_{\rm tot}$ are the generators of the SO(4) symmetry group of the half-filled Kondo model on the bipartite lattice \cite{TSU97b} -- and of the effective model as well.
The effective isospin interaction is ``antiferromagnetic''.
Analogous to the Mermin-Wagner theorem \cite{MW66}, and opposed to ferromagnetic spin order, antiferromagnetic (staggered) isospin order would be suppressed by quantum fluctuations of the order parameter for $D=1$.
The necessary formation of local isospin moments in the ground state is suppressed anyway by the repulsive Hubbard term (third term in Eq.\ (\ref{eq:eff})). 
On the contrary, the Hubbard interaction favors formation of local {\em magnetic} moments.
Finally, there is a correlated hopping term in $H_{\rm eff}$ which, however, is only active between a spin at $i$ and an isospin at $j$ or vice versa. 
Exact diagonalization of $H_{\rm eff}$ for systems with a few A-sites is easily done and in fact yields a ferromagnetic ground state with $S_{\rm tot,0}=(R-1)/2$ while $\langle \ff t_{i}^{2} \rangle \equiv 0$.

The effective model and thus the IIME concept is also valid for fillings $n\ne 1$, as long as the local Kondo singlets in the $t=0$ ground state are unbroken, i.e.\ for fillings $1/2 \le n \le 3/2$.
In dimensions $D>1$ essentially the same effective model is obtained. 
Generalizations to non-bipartite lattice structures are possible.

%--------------------------------------------------------------------------------------------------------------------------------------
\paragraph{Charge fluctuations.}

The IIME mechanism is robust against charge fluctuations on the impurities. 
This is demonstrated by DMRG calculations where the spin-1/2 Kondo impurities are replaced by
Anderson impurities, i.e., correlated sites with Hubbard interaction $U$ coupled to the conduction electrons by a local hybridization $V$.
For weak $V$ with $8V^{2}/U = J \ll t$ the results for the two models agree (filled and open symbols in Fig.\ \ref{fig:mag}), as prescribed by the Schrieffer-Wolff transformation \cite{SW66}. 
With increasing $V$, however, we again find a crossover from RKKY to IIME, and for strong $V$ confinement of A-site electrons is due to the formation of strongly bound states at the $B$ sites.
In the case of Kondo impurities the crossover takes place between $J/t =2$ and $J/t = 4$ while in the Anderson case it is located around $V/t = 2$.

%--------------------------------------------------------------------------------------------------------------------------------------
\begin{figure}[t]
\centerline{\includegraphics[width=0.45\textwidth,angle=0]{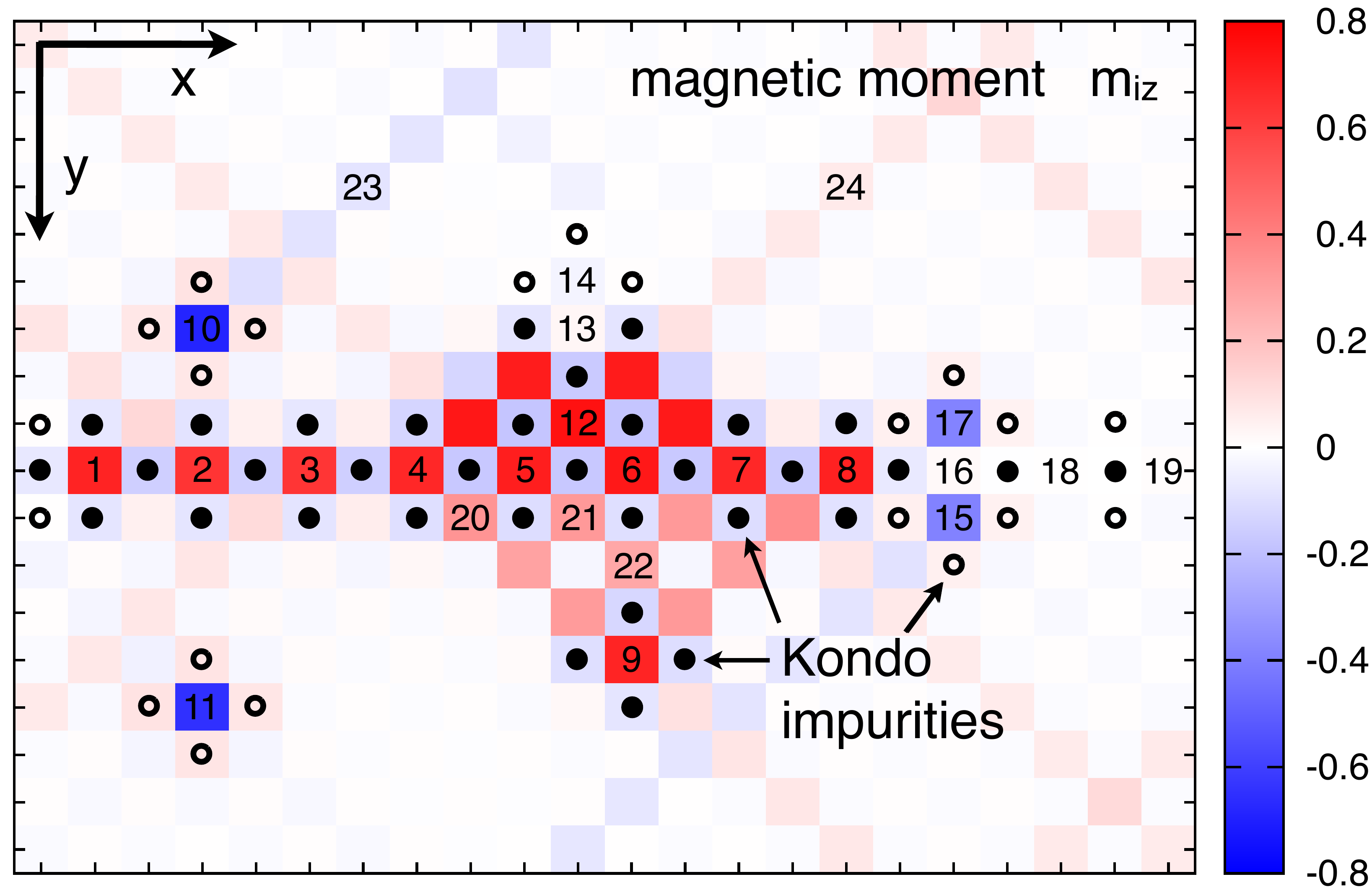}}
\caption{(Color online) 
Magnetic structure of an IIME-coupled system in an artificial $D=2$ geometry.
$R=57$ Kondo impurities (filled and open dots) are placed on an $L=22\times 18$ array of uncorrelated sites and coupled locally ($J=5$). 
Real-space DMFT results for the ordered magnetic moments (color code) at sites in the uncorrelated layer.
Periodic boundary conditions are used. Labels: see text.
}
\label{fig:2d}
\end{figure}
%--------------------------------------------------------------------------------------------------------------------------------------

%--------------------------------------------------------------------------------------------------------------------------------------
\paragraph{Dynamical mean-field theory.}

While non-local correlations due to the RKKY interaction in the metastable paramagnetic state are not accessible to a mean-field description, at least for weak $V$ \cite{TSRP12}, DMFT is able, however, to describe the symmetry-broken ferromagnetic state.
The solid and dashed lines in Fig.\ \ref{fig:mag} show the results of DMFT calculations for Kondo and Anderson impurities, respectively. 
We employ a standard Lanczos implementation \cite{TSRP12,CK94}. 
In case of the diluted Kondo-lattice model, Eq.\ (\ref{eq:ham}), the effective self-consistently determined impurity problem consists of the local spin $\ff S_{r}$, the corresponding B site and up to 8 bath sites \cite{NMK00,OKK09a}, while up to 9 bath sites are used in the Anderson case.
Almost perfect agreement with the DMRG data is found, see Fig.\ \ref{fig:mag}. 

DMFT can be employed to describe the magnetic properties of nanostructures of magnetic atoms on a $D=2$ metallic surface layer (isolated from the substrate by a spacer).
Motivated by the agreement with DMRG for the symmetry-broken state of the $D=1$ bulk system, we again expect quantitatively reliable predictions.
This can be checked to some extent in case of a bipartite structure as the total spin quantum number is fixed by reflection positivity in spin space \cite{She96,Tsu97a}.
More important from a physical point of view, however, is to confine, with the help of the Kondo effect, conduction electrons in certain geometries to avoid a dissipation of the total spin into a large (bulk) layer.

Fig.\ \ref{fig:2d} displays an example for a $D=2$ layer with several magnetic impurities. 
Calculations are done using the real-space generalization of DMFT \cite{PN97b} for the corresponding diluted Kondo lattice model. 
A symmetry-broken ground state is found:
Electrons in the chain of A-sites 1, 2, ..., 8 with relative distance $\delta =2$ are confined.
Their local moments in fact order ferromagnetically. 
The moment at A-site 9 (distance $\delta = 4$) couples ferromagnetically while the moments at sites 10 and 11 ($\delta=3$ and $\delta = 5$) couple antiferromagnetically to the chain. 
Note that the local moments at 10 and 11 are formed by confinement due to surrounding local Kondo singlets.
If such moments are weakly coupled to the rest of the system, a tiny Weiss field produced by the chain is sufficient to result in an almost full polarization.
$|m_{\rm B}|$ is in fact found to slightly increase with increasing distance from the chain. 
We conclude that the IIME is oscillatory and decreasing with distance -- similar to the RKKY case.

More characteristic features of the IIME can be studied qualitatively (see Fig.\ \ref{fig:2d}):
Neighboring A-sites with higher effective coordination mutually support magnetic polarization.
This explains the slightly enhanced $m_{\rm A}$ at and around A-site 12.
Confinement of an {\em odd} number of electrons is important: 
There is almost no moment at sites 13 and 14 while the structure 15, 16, 17 is polarized.
Confinement as such (with respect to all dimensions) is essential: 
Electrons at 18 or 19 are not confined, no local moments are formed and thus no polarization found.
There is a proximity effect, however, as can be seen at 20, 21, or 22.
Furthermore, spin-dependent multiple scattering of conduction electrons at the magnetic structures causes an interference pattern,
see sites 23, 24, for example.
The real-space DMFT is found to give {\em quantitatively} reliable results:
Summing up the local magnetic moments for the 396 uncorrelated sites and the 57 impurities, we find a ground-state spin moment of $m_{\rm tot} = 2 S_{\rm tot, z} = 15.13$ which is, within numerical uncertainties, equal to the {\rm exact} value $m_{\rm tot} = 36 - 21= 15$ which can be obtained analytically \cite{She96,Tsu97a,TSU97b} by counting the number of impurities on B (filled) and on A sites (open dots).

%--------------------------------------------------------------------------------------------------------------------------------------
\paragraph{Conclusions.}

We have analyzed an indirect magnetic exchange mechanism where confinement of conduction electrons due to scattering at Kondo singlets leads to local-moment formation at {\em a priori} uncorrelated sites and to spin and isospin couplings via virtual excitations of the Kondo singlets.
This IIME is ``inverse'' to the conventional RKKY coupling. 
Its oscillatory distance dependence can be utilized to construct nanostructures with tailored magnetic properties, e.g., by placing magnetic atoms in certain geometries on a metallic layer, similar to RKKY-based artificial structures \cite{ZWL+10,KWCW11,KWC+12}.
Alternatively, systems of ultracold fermionic atoms trapped in optical lattices
\cite{JZ05,BDN12}
may realize multi-impurity Kondo systems in the strong-coupling regime essential to the IIME.
Future theoretical work may explore systems with correlated conduction electrons and preformed local moments. 
Spin-only, e.g., Kondo necklace models may be considered to get an {\em analytical} expression for the effective IIME spin coupling. 
%IIME in ferromagnetic ($J<0$) multi-impurity Kondo models and in spin-only, e.g., Kondo necklace models. 
DMFT and DMRG studies of filling dependencies appear particularly exciting. 
Temperature dependencies are accessible to quantum Monte-Carlo techniques \cite{Ass02} on bipartite lattices at half-filling.

\acknowledgments
We thank D.\ G\"utersloh for instructive discussions.
Support of this work by the Deutsche Forschungsgemeinschaft within the SFB 668 (project A14), by the excellence cluster ``The Hamburg Centre for Ultrafast Imaging - Structure, Dynamics and Control of Matter at the Atomic Scale'' and by the SFB 925 (project B5) is gratefully acknowledged.


\begin{thebibliography}{36}
\expandafter\ifx\csname natexlab\endcsname\relax\def\natexlab#1{#1}\fi
\expandafter\ifx\csname bibnamefont\endcsname\relax
  \def\bibnamefont#1{#1}\fi
\expandafter\ifx\csname bibfnamefont\endcsname\relax
  \def\bibfnamefont#1{#1}\fi
\expandafter\ifx\csname citenamefont\endcsname\relax
  \def\citenamefont#1{#1}\fi
\expandafter\ifx\csname url\endcsname\relax
  \def\url#1{\texttt{#1}}\fi
\expandafter\ifx\csname urlprefix\endcsname\relax\def\urlprefix{URL }\fi
\providecommand{\bibinfo}[2]{#2}
\providecommand{\eprint}[2][]{\url{#2}}

\bibitem[{\citenamefont{Yosida}(1996)}]{Yos96}
\bibinfo{author}{\bibfnamefont{K.}~\bibnamefont{Yosida}},
  \emph{\bibinfo{title}{Theory of Magnetism}}, vol. \bibinfo{volume}{122} of
  \emph{\bibinfo{series}{Springer Series in Solid-State Physics}}
  (\bibinfo{publisher}{Springer}, \bibinfo{address}{Berlin},
  \bibinfo{year}{1996}).

\bibitem[{\citenamefont{Nolting and Ramakanth}(2009)}]{NR09}
\bibinfo{author}{\bibfnamefont{W.}~\bibnamefont{Nolting}} \bibnamefont{and}
  \bibinfo{author}{\bibfnamefont{A.}~\bibnamefont{Ramakanth}},
  \emph{\bibinfo{title}{Quantum Theory of Magnetism}}
  (\bibinfo{publisher}{Springer}, \bibinfo{address}{Berlin},
  \bibinfo{year}{2009}).

\bibitem[{\citenamefont{Anderson}(1961)}]{And61}
\bibinfo{author}{\bibfnamefont{P.~W.} \bibnamefont{Anderson}},
  \bibinfo{journal}{Phys. Rev.} \textbf{\bibinfo{volume}{124}},
  \bibinfo{pages}{41} (\bibinfo{year}{1961}).

\bibitem[{\citenamefont{Hubbard}(1963)}]{Hub63}
\bibinfo{author}{\bibfnamefont{J.}~\bibnamefont{Hubbard}},
  \bibinfo{journal}{Proc. R. Soc. London A} \textbf{\bibinfo{volume}{276}},
  \bibinfo{pages}{238} (\bibinfo{year}{1963}).

\bibitem[{\citenamefont{Kondo}(1964)}]{Kon64}
\bibinfo{author}{\bibfnamefont{J.}~\bibnamefont{Kondo}},
  \textbf{\bibinfo{volume}{32}}, \bibinfo{pages}{37} (\bibinfo{year}{1964}).

\bibitem[{\citenamefont{Heisenberg}(1928)}]{hei28}
\bibinfo{author}{\bibfnamefont{W.~J.} \bibnamefont{Heisenberg}},
  \bibinfo{journal}{Z. Phys.} \textbf{\bibinfo{volume}{49}},
  \bibinfo{pages}{619} (\bibinfo{year}{1928}).

\bibitem[{\citenamefont{Kramers}(1934)}]{Kra34}
\bibinfo{author}{\bibfnamefont{H.~A.} \bibnamefont{Kramers}},
  \bibinfo{journal}{Physica} \textbf{\bibinfo{volume}{1}}, \bibinfo{pages}{182}
  (\bibinfo{year}{1934}).

\bibitem[{\citenamefont{Anderson}(1950)}]{And50}
\bibinfo{author}{\bibfnamefont{P.~W.} \bibnamefont{Anderson}},
  \bibinfo{journal}{Phys. Rev.} \textbf{\bibinfo{volume}{79}},
  \bibinfo{pages}{350} (\bibinfo{year}{1950}).

\bibitem[{\citenamefont{Ruderman and Kittel}(1954)}]{RK54}
\bibinfo{author}{\bibfnamefont{M.~A.} \bibnamefont{Ruderman}} \bibnamefont{and}
  \bibinfo{author}{\bibfnamefont{C.}~\bibnamefont{Kittel}},
  \bibinfo{journal}{Phys. Rev.} \textbf{\bibinfo{volume}{96}},
  \bibinfo{pages}{99} (\bibinfo{year}{1954}).

\bibitem[{\citenamefont{Kasuya}(1956)}]{Kas56}
\bibinfo{author}{\bibfnamefont{T.}~\bibnamefont{Kasuya}},
  \bibinfo{journal}{Prog. Theor. Phys.} \textbf{\bibinfo{volume}{16}},
  \bibinfo{pages}{45} (\bibinfo{year}{1956}).

\bibitem[{\citenamefont{Yosida}(1957)}]{Yos57}
\bibinfo{author}{\bibfnamefont{K.}~\bibnamefont{Yosida}},
  \bibinfo{journal}{Phys. Rev.} \textbf{\bibinfo{volume}{106}},
  \bibinfo{pages}{893} (\bibinfo{year}{1957}).

\bibitem[{\citenamefont{Craig et~al.}(2004)\citenamefont{Craig, Taylor, Lester,
  Marcus, Hanson, and Gossard}}]{CTL+04}
\bibinfo{author}{\bibfnamefont{N.~J.} \bibnamefont{Craig}},
  \bibinfo{author}{\bibfnamefont{J.~M.} \bibnamefont{Taylor}},
  \bibinfo{author}{\bibfnamefont{E.~A.} \bibnamefont{Lester}},
  \bibinfo{author}{\bibfnamefont{C.~M.} \bibnamefont{Marcus}},
  \bibinfo{author}{\bibfnamefont{M.~P.} \bibnamefont{Hanson}},
  \bibnamefont{and} \bibinfo{author}{\bibfnamefont{A.~C.}
  \bibnamefont{Gossard}}, \bibinfo{journal}{Science}
  \textbf{\bibinfo{volume}{304}}, \bibinfo{pages}{565} (\bibinfo{year}{2004}).

\bibitem[{\citenamefont{Zhou et~al.}(2010)\citenamefont{Zhou, Wiebe, Lounis,
  Vedmedenko, Meier, Bl\"ugel, Dederichs, and Wiesendanger}}]{ZWL+10}
\bibinfo{author}{\bibfnamefont{L.}~\bibnamefont{Zhou}},
  \bibinfo{author}{\bibfnamefont{J.}~\bibnamefont{Wiebe}},
  \bibinfo{author}{\bibfnamefont{S.}~\bibnamefont{Lounis}},
  \bibinfo{author}{\bibfnamefont{E.}~\bibnamefont{Vedmedenko}},
  \bibinfo{author}{\bibfnamefont{F.}~\bibnamefont{Meier}},
  \bibinfo{author}{\bibfnamefont{S.}~\bibnamefont{Bl\"ugel}},
  \bibinfo{author}{\bibfnamefont{P.}~\bibnamefont{Dederichs}},
  \bibnamefont{and}
  \bibinfo{author}{\bibfnamefont{R.}~\bibnamefont{Wiesendanger}},
  \bibinfo{journal}{Nature Physics} \textbf{\bibinfo{volume}{6}},
  \bibinfo{pages}{187} (\bibinfo{year}{2010}).

\bibitem[{\citenamefont{Khajetoorians et~al.}(2011)\citenamefont{Khajetoorians,
  Wiebe, Chilian, and Wiesendanger}}]{KWCW11}
\bibinfo{author}{\bibfnamefont{A.~A.} \bibnamefont{Khajetoorians}},
  \bibinfo{author}{\bibfnamefont{J.}~\bibnamefont{Wiebe}},
  \bibinfo{author}{\bibfnamefont{B.}~\bibnamefont{Chilian}}, \bibnamefont{and}
  \bibinfo{author}{\bibfnamefont{R.}~\bibnamefont{Wiesendanger}},
  \bibinfo{journal}{Science} \textbf{\bibinfo{volume}{332}},
  \bibinfo{pages}{1062} (\bibinfo{year}{2011}).

\bibitem[{\citenamefont{Khajetoorians et~al.}(2012)\citenamefont{Khajetoorians,
  Wiebe, Chilian, Lounis, Bl\"ugel, and Wiesendanger}}]{KWC+12}
\bibinfo{author}{\bibfnamefont{A.~A.} \bibnamefont{Khajetoorians}},
  \bibinfo{author}{\bibfnamefont{J.}~\bibnamefont{Wiebe}},
  \bibinfo{author}{\bibfnamefont{B.}~\bibnamefont{Chilian}},
  \bibinfo{author}{\bibfnamefont{S.}~\bibnamefont{Lounis}},
  \bibinfo{author}{\bibfnamefont{S.}~\bibnamefont{Bl\"ugel}}, \bibnamefont{and}
  \bibinfo{author}{\bibfnamefont{R.}~\bibnamefont{Wiesendanger}},
  \bibinfo{journal}{Nature Physics} \textbf{\bibinfo{volume}{8}},
  \bibinfo{pages}{497} (\bibinfo{year}{2012}).

\bibitem[{\citenamefont{Doniach}(1977)}]{Don77}
\bibinfo{author}{\bibfnamefont{S.}~\bibnamefont{Doniach}},
  \bibinfo{journal}{Physica B} \textbf{\bibinfo{volume}{91}},
  \bibinfo{pages}{321} (\bibinfo{year}{1977}).

\bibitem[{\citenamefont{Hewson}(1993)}]{hew93}
\bibinfo{author}{\bibfnamefont{A.~C.} \bibnamefont{Hewson}},
  \emph{\bibinfo{title}{The Kondo Problem to Heavy Fermions}}
  (\bibinfo{publisher}{Cambridge University Press},
  \bibinfo{address}{Cambridge}, \bibinfo{year}{1993}).

\bibitem[{\citenamefont{Thimm et~al.}(1999)\citenamefont{Thimm, Kroha, and von
  Delft}}]{TKvD99}
\bibinfo{author}{\bibfnamefont{W.}~\bibnamefont{Thimm}},
  \bibinfo{author}{\bibfnamefont{J.}~\bibnamefont{Kroha}}, \bibnamefont{and}
  \bibinfo{author}{\bibfnamefont{J.}~\bibnamefont{von Delft}},
  \bibinfo{journal}{Phys. Rev. Lett.} \textbf{\bibinfo{volume}{82}},
  \bibinfo{pages}{2143} (\bibinfo{year}{1999}).

\bibitem[{\citenamefont{Schwabe et~al.}(2012)\citenamefont{Schwabe,
  G\"utersloh, and Potthoff}}]{SGP12}
\bibinfo{author}{\bibfnamefont{A.}~\bibnamefont{Schwabe}},
  \bibinfo{author}{\bibfnamefont{D.}~\bibnamefont{G\"utersloh}},
  \bibnamefont{and} \bibinfo{author}{\bibfnamefont{M.}~\bibnamefont{Potthoff}},
  \bibinfo{journal}{Phys. Rev. Lett.} \textbf{\bibinfo{volume}{109}},
  \bibinfo{pages}{257202} (\bibinfo{year}{2012}).

\bibitem[{\citenamefont{Shen}(1996)}]{She96}
\bibinfo{author}{\bibfnamefont{S.-Q.} \bibnamefont{Shen}},
  \bibinfo{journal}{Phys. Rev. B} \textbf{\bibinfo{volume}{53}},
  \bibinfo{pages}{14252} (\bibinfo{year}{1996}).

\bibitem[{\citenamefont{Tsunetsugu}(1997)}]{Tsu97a}
\bibinfo{author}{\bibfnamefont{H.}~\bibnamefont{Tsunetsugu}},
  \bibinfo{journal}{Phys. Rev. B} \textbf{\bibinfo{volume}{55}},
  \bibinfo{pages}{3042} (\bibinfo{year}{1997}).

\bibitem[{\citenamefont{Lieb}(1989)}]{Lie89}
\bibinfo{author}{\bibfnamefont{E.~H.} \bibnamefont{Lieb}},
  \bibinfo{journal}{Phys. Rev. Lett.} \textbf{\bibinfo{volume}{62}},
  \bibinfo{pages}{1201} (\bibinfo{year}{1989}).

\bibitem[{\citenamefont{White}(1992)}]{Whi92}
\bibinfo{author}{\bibfnamefont{S.~R.} \bibnamefont{White}},
  \bibinfo{journal}{Phys. Rev. Lett.} \textbf{\bibinfo{volume}{69}},
  \bibinfo{pages}{2863} (\bibinfo{year}{1992}).

\bibitem[{\citenamefont{Schollw\"ock}(2011)}]{Sch11}
\bibinfo{author}{\bibfnamefont{U.}~\bibnamefont{Schollw\"ock}},
  \bibinfo{journal}{Ann. Phys. (N.Y.)} \textbf{\bibinfo{volume}{326}},
  \bibinfo{pages}{96} (\bibinfo{year}{2011}).

\bibitem[{\citenamefont{Titvinidze et~al.}(2012)\citenamefont{Titvinidze,
  Schwabe, Rother, and Potthoff}}]{TSRP12}
\bibinfo{author}{\bibfnamefont{I.}~\bibnamefont{Titvinidze}},
  \bibinfo{author}{\bibfnamefont{A.}~\bibnamefont{Schwabe}},
  \bibinfo{author}{\bibfnamefont{N.}~\bibnamefont{Rother}}, \bibnamefont{and}
  \bibinfo{author}{\bibfnamefont{M.}~\bibnamefont{Potthoff}},
  \bibinfo{journal}{Phys. Rev. B} \textbf{\bibinfo{volume}{86}},
  \bibinfo{pages}{075141} (\bibinfo{year}{2012}).

\bibitem[{\citenamefont{Titvinidze et~al.}(2013)\citenamefont{Titvinidze,
  Schwabe, and Potthoff}}]{TSP13}
\bibinfo{author}{\bibfnamefont{I.}~\bibnamefont{Titvinidze}},
  \bibinfo{author}{\bibfnamefont{A.}~\bibnamefont{Schwabe}}, \bibnamefont{and}
  \bibinfo{author}{\bibfnamefont{M.}~\bibnamefont{Potthoff}},
  \bibinfo{journal}{unpublished}  (\bibinfo{year}{2013}).

\bibitem[]{TDH+13}
A. C. Tiegel, P. E. Dargel, K. A. Hallberg, H. Frahm and T. Pruschke, 
Phys. Rev. B \textbf{87}, 075122 (2013).

\bibitem[{\citenamefont{Mermin and Wagner}(1966)}]{MW66}
\bibinfo{author}{\bibfnamefont{N.~D.} \bibnamefont{Mermin}} \bibnamefont{and}
  \bibinfo{author}{\bibfnamefont{H.}~\bibnamefont{Wagner}},
  \bibinfo{journal}{Phys. Rev. Lett.} \textbf{\bibinfo{volume}{17}},
  \bibinfo{pages}{1133} (\bibinfo{year}{1966}).

\bibitem[{\citenamefont{Schrieffer and Wolff}(1966)}]{SW66}
\bibinfo{author}{\bibfnamefont{J.~R.} \bibnamefont{Schrieffer}}
  \bibnamefont{and} \bibinfo{author}{\bibfnamefont{P.~A.} \bibnamefont{Wolff}},
  \bibinfo{journal}{Phys. Rev.} \textbf{\bibinfo{volume}{149}},
  \bibinfo{pages}{491} (\bibinfo{year}{1966}).

\bibitem[{\citenamefont{Caffarel and Krauth}(1994)}]{CK94}
\bibinfo{author}{\bibfnamefont{M.}~\bibnamefont{Caffarel}} \bibnamefont{and}
  \bibinfo{author}{\bibfnamefont{W.}~\bibnamefont{Krauth}},
  \bibinfo{journal}{Phys. Rev. Lett.} \textbf{\bibinfo{volume}{72}},
  \bibinfo{pages}{1545} (\bibinfo{year}{1994}).

\bibitem[{\citenamefont{Momoi and Kubo}(2000)}]{NMK00}
\bibinfo{author}{\bibfnamefont{K.~N.~T.} \bibnamefont{Momoi}} \bibnamefont{and}
  \bibinfo{author}{\bibfnamefont{K.}~\bibnamefont{Kubo}}, \bibinfo{journal}{J.
  Phys. Soc. Jpn.} \textbf{\bibinfo{volume}{69}}, \bibinfo{pages}{1837}
  (\bibinfo{year}{2000}).

\bibitem[{\citenamefont{Otsuki et~al.}(2009)\citenamefont{Otsuki, Kusunose, and
  Kuramoto}}]{OKK09a}
\bibinfo{author}{\bibfnamefont{J.}~\bibnamefont{Otsuki}},
  \bibinfo{author}{\bibfnamefont{H.}~\bibnamefont{Kusunose}}, \bibnamefont{and}
  \bibinfo{author}{\bibfnamefont{Y.}~\bibnamefont{Kuramoto}},
  \bibinfo{journal}{J. Phys. Soc. Jpn.} \textbf{\bibinfo{volume}{78}},
  \bibinfo{pages}{014702} (\bibinfo{year}{2009}).

\bibitem[{\citenamefont{Potthoff and Nolting}(1997)}]{PN97b}
\bibinfo{author}{\bibfnamefont{M.}~\bibnamefont{Potthoff}} \bibnamefont{and}
  \bibinfo{author}{\bibfnamefont{W.}~\bibnamefont{Nolting}},
  \bibinfo{journal}{Phys. Rev. B} \textbf{\bibinfo{volume}{55}},
  \bibinfo{pages}{2741} (\bibinfo{year}{1997}).
\bibitem[{\citenamefont{Tsunetsugu et~al.}(1997)\citenamefont{Tsunetsugu,
  Sigrist, and Ueda}}]{TSU97b}

\bibinfo{author}{\bibfnamefont{H.}~\bibnamefont{Tsunetsugu}},
  \bibinfo{author}{\bibfnamefont{M.}~\bibnamefont{Sigrist}}, \bibnamefont{and}
  \bibinfo{author}{\bibfnamefont{K.}~\bibnamefont{Ueda}},
  \bibinfo{journal}{Rev. Mod. Phys.} \textbf{\bibinfo{volume}{69}},
  \bibinfo{pages}{809} (\bibinfo{year}{1997}).

\bibitem[{\citenamefont{Jaksch and Zoller}(2005)}]{JZ05}
\bibinfo{author}{\bibfnamefont{D.}~\bibnamefont{Jaksch}} \bibnamefont{and}
  \bibinfo{author}{\bibfnamefont{P.}~\bibnamefont{Zoller}},
  \bibinfo{journal}{Ann. Phys. (N.Y.)} \textbf{\bibinfo{volume}{315}},
  \bibinfo{pages}{52} (\bibinfo{year}{2005}).

\bibitem[{\citenamefont{Bloch et~al.}(2012)\citenamefont{Bloch, Dalibard, and
  Nascimb\`e{}ne}}]{BDN12}
\bibinfo{author}{\bibfnamefont{I.}~\bibnamefont{Bloch}},
  \bibinfo{author}{\bibfnamefont{J.}~\bibnamefont{Dalibard}}, \bibnamefont{and}
  \bibinfo{author}{\bibfnamefont{S.}~\bibnamefont{Nascimb\`e{}ne}},
  \bibinfo{journal}{Nature Physics} \textbf{\bibinfo{volume}{8}},
  \bibinfo{pages}{267} (\bibinfo{year}{2012}).

\bibitem[{\citenamefont{Assaad}(2002)}]{Ass02}
\bibinfo{author}{\bibfnamefont{F.~F.} \bibnamefont{Assaad}},
  \bibinfo{journal}{Phys. Rev. B} \textbf{\bibinfo{volume}{65}},
  \bibinfo{pages}{115104} (\bibinfo{year}{2002}).

\end{thebibliography}
\end{document}